\documentclass[runningheads]{llncs}
\usepackage{graphicx}
\usepackage{xcolor}
\usepackage{algorithmic}
\usepackage{algorithm}
\usepackage{caption}
\usepackage{subcaption}
\usepackage{wrapfig,lipsum,booktabs}
\usepackage{makecell}
\begin{document}
%
\title{Unveiling and Mitigating Bias in Ride-Hailing Pricing for Equitable Policy Making$^*$}
%
\titlerunning{Unveiling and Mitigating Bias in Ride-Hailing Pricing}
%
\author{Submitted for blind review \inst{ } } 
\institute{}
\author{Nripsuta Ani Saxena $^{**,}$\inst{1}\and
Wenbin Zhang \inst{2} \and
Cyrus Shahabi \inst{1}}
\authorrunning{N. Saxena et al.}
%
\institute{University of Southern California, Los Angeles, CA 90089, USA \and Michigan Technological University, Houghton, MI 49931, USA
}


%
%
\maketitle              
\begin{abstract}

Ride-hailing services have skyrocketed in popularity due to the convenience they offer, but recent research has shown that their pricing strategies can have a disparate impact on some riders, such as those living in disadvantaged neighborhoods with a greater share of residents of color or residents below the poverty line. Since these communities tend to be more dependent on ride-hailing services due to lack of adequate public transportation, it is imperative to address this inequity. 
To this end, this paper presents the first thorough study on fair pricing for ride-hailing services by devising applicable fairness measures and corresponding fair pricing mechanisms. By providing discounts that may be subsidized by the government, our approach results in an increased number and more affordable rides for the disadvantaged community. 
Experiments on real-world Chicago taxi data confirm our theoretical findings which provide a basis for the government to establish fair ride-hailing policies.

\end{abstract}
\let\thefootnote\relax\footnotetext{$^*$ Preconference version: Not final.}
\let\thefootnote\relax\footnotetext{$^{**}$ Correspondence should be directed to: nsaxena@usc.edu.}

\section{Introduction}


Since Uber first launched in 2009, ride-hailing companies have evolved into a global service that has become an indelible aspect of our lives~\cite{willis2021using,wallsten2015competitive}. While such services have had positive effects, providing consumers with more choices and comfort, some aspects of these services might have negative social effects \cite{pandey2021disparate,ge2016racial,chang2021does,zhang2019faht}. For example, research has shown that the pricing algorithms of ride-hailing services can lead to disparate impact: neighborhoods with a higher proportion of people of color, higher levels of poverty, and younger residents are significantly associated with higher fare prices \cite{pandey2021disparate}. This disparate impact is especially troubling because it further harms communities who tend to be more dependent on these services in the first place due to poor public transportation connectivity and who have historically been systemically marginalized in a myriad of ways.  


City governments all over the country already recognize a disparity in transportation options available to its disadvantaged residents, and ride-hailing apps worsen this inequity. The City of Chicago has even made reducing inequities in mobility for all its residents one of the five essential elements of its development plan \textit{On To 2050} \cite{OnTo2050} which focuses on inclusive growth. Many local governments are enacting policies to help address this injustice \cite{AffCon,LABus,USDoT,LAMetro}. Unfortunately, none of these policies help alleviate the additional disparate impact caused by the pricing mechanisms of ride-hailing companies. While they are a step in the right direction, these policies are not enough on their own since disadvantaged neighborhoods do not have enough efficient public transportation coverage \cite{garrett1999reconsidering,kramer2018unaffordable,sanchez2008poverty,zhang2022longitudinal}, which leads to greater dependence on ride-hailing \cite{cats2022beyond}. Indeed, researchers analyzing Uber trip data for six large cities in the U.S.A. and Europe found that 20 to 40\% of ride-hailing trips had no viable public transportation alternative available \cite{cats2022beyond}. This work aims to provide a technique for the government to help address the worsened inequity. Discounts proposed by our pricing mechanisms can be covered by the government as subsidies to reduce disparate impact for disadvantaged neighborhoods, in the same vein as other subsidy initiatives for low-income residents for necessary services \cite{AffCon,LABus,USDoT,LAMetro}. 

Although research into pricing mechanisms for ride-hailing services is a burgeoning area, most work looks at improving the efficiency of matching drivers and riders and pricing rides at times of low supply \cite{tong2018dynamic}, or optimizing pricing to regulate demand on the platform \cite{yuan2021real}. There is little work examining fair pricing to reduce disparate impact. Relevantly, fares for Uber rides in USA from major airports to hotels were found to be significantly correlated with the prices of hotel rooms in \cite{chang2021does}, although they do not attempt to address this price discrimination in ride-hailing. Fair benefits for drivers on the ride-hailing platform has also been investigated, but fair prices for the riders are not considered \cite{suhr2019two}. In \cite{lu2020say}, the hypothetical scenario where ride-hailing services use personal data about customers to determine customized prices for each customer is studied. While this can help address a form of discrimination, it does not explicitly address the disparate impact caused by current pricing techniques.

 


Despite its importance and necessity, unveiling and mitigating the disparate impact of ride-hailing apps' pricing mechanisms, with the goal of providing a foundation for the government for drafting fair ride-hailing policy presents unique challenges. First, there is a difference in affordability. Different neighborhoods may have different average levels of income, and thus what may be affordable for a resident of one neighborhood may not be affordable for the resident of another. A fair, equitable solution should price rides according to affordability so that no one gets priced out. Second, there are data-related challenges since all the data ideally needed to model the problem holistically is unavailable. For example, metrics such as price elasticity, which in our case is the measure of change in the number of trips with a change in the price of trips, is unknown, as is the surge level of trips, and thus these quantities must be estimated, and require making assumptions at times. 

To address the aforementioned challenges, this paper introduces novel definitions of fair pricing and corresponding fair-pricing strategies to address discrimination in ride-hailing. More specifically, the main contributions of this paper are: i) We use the concepts of price elasticity and consumer surplus from the economics literature to explore the affordability of trips for different groups and the relationship between pricing and the number of rides that take place to formally define bias in ride-hailing. ii) We propose a fair-pricing strategy that effectively reduces disparate impact and is versatile in accommodating different requirements for multiple scenarios. iii) We use real-world ride-hailing data from Chicago and census data to conduct several experiments comparing our pricing mechanisms to other models commonly used for pricing, validating the real-world utility of our method.

\section{Bias in Ride-Hailing}


Biased pricing in ride-hailing results in higher prices for people who are likely to already be struggling financially. To address this disparate impact due to the ride-hailing services' black-box pricing algorithms, we consider bias in ride-hailing to be a significant difference in the average rides taken by the disadvantaged versus those of non-disadvantaged groups and corresponding affordability.

To formally define disadvantaged and non-disadvantaged groups, we use the classification by the Chicago Metropolitan Agency for Planning (CMAP) which identifies census tracts in the Chicago region, called \textit{Economically Disconnected Areas} (EDAs) that are disconnected from economic growth and prosperity, and may be experiencing disinvestment \cite{WinNT}. In other words, EDAs are neighborhoods with high concentrations of low-income households and minorities or households with limited English proficiency speakers. Approximately one-third of the city lives in an EDA \cite{WinNT}. In the remainder of this paper, we interchangeably use the terms \textbf{EDA} regions and disadvantaged neighborhoods as well as \textbf{non-EDA} regions and non-disadvantaged neighborhoods for readability. We further define \textbf{EDA-trips} to denote trips (or rides) that either begin or end in an EDA region. \textbf{Non-EDA-trips} denotes trips (or rides) that do not. 

\section{Unveiling Ride-Hailing Bias}
\subsection{Relative Rideability}


Aligned with the $p\%$-rule \cite{biddle2017adverse} used by the U.S. Equal Employment Opportunity Commission (EEOC) to evaluate disparate impact, we introduce the \textit{Relative Rideability} ($R^2$) score to quantify the previously discussed bias in ride-hailing shown as the significant difference in the average rides across different groups. Specifically, the $p\%$-rule states that if the selection rate for a certain group is less than $p\%$ of the selection rate for the group with the highest selection rate, then there is a substantially different rate of selection and may be considered disparate impact. Analogously, $R^2$ can be mathematically defined as below: 


\begin{equation}
R^2 = \frac{\min\{d_1,\dots, d_i\}}{\max\{\neg d_1,\dots, \neg d_j\}}
\end{equation} 

\noindent where $d_{i}$ is the average number of trips by residents of disadvantaged group $i$, while $\neg d_{j}$ is the average number of trips by residents of non-disadvantaged group $j$.
With $\neg d_{j}$ commonly greater than $d_{i}$ in inequitable ride-hailing services, the higher the $R^2$ the fairer the model. \\

\subsection{Affordability}
 \label{DI}

Next we look at another way of quantifying bias in ride-hailing: difference in ride affordability. A measure of affordability can be captured by consumer surplus \cite{xiao2020effects},
a concept from economics, that captures the difference between a consumer's willingness to pay for a certain product or service, and the price of that product or service. If the price of the product or service is less than the amount the customer is willing to pay for it, then the consumer surplus is positive. Otherwise, it is negative, which can happen if the good or service is necessary (e.g., food, or life-saving medication).  In other words, the higher the consumer surplus, the more easily the consumer can afford that product or service. 




However, the information needed to quantify consumer surplus in ride-hailing, such as the number of people who looked at the quoted price but did not make a request and what that price was, is unavailable in publicly released ride-hailing data. We circumvent this challenge by computing \textit{price elasticity ($E_p$)} instead, which measures the change in demand of a product or service with respect to a change in its price, to obtain an estimation of consumer surplus:


\begin{equation} 
\label{price_elas}
E_{p} = \frac{\delta_Q}{\delta_P} 
\end{equation} 

\vspace{0.14cm}
\noindent where $\delta_Q$ is the percentage change in the quantity of the product or service demanded, and $\delta_P$ is the percentage change in its price. As a note, price elasticity will also be used in our following fair pricing mechanisms to estimate how the number of trips might change with a change in price. 

Price elasticity is considered at a point where there is a change in the price of a good or service. However, ride-hailing trips are not assigned a flat per-mile rate, thus simply studying trips at different prices is not ideal since the difference in price of two trips might be caused by multiple factors. We therefore study similar trips that were shown different prices due to the way ride-hailing platforms compute surge level, a multiplier that is used to multiply and increase the estimated price of a ride at the time of high demand or low supply. In addition, ride-hailing platforms typically compute a continuous surge level for rides, but show a discrete value to consumers for simplicity and ease of understanding. For example, a trip for which a surge level was computed as 1.449 will result in a discretized surge of 1.4x, whereas a surge level of 1.451x will result in a discretized surge of 1.5x. We make use of regression discontinuity design around these discretized surge points to estimate price elasticities when surge levels go from 1.2x to 1.3x, 1.3x to 1.4x, 1.4x to 1.5x, and so forth. We run a linear probability model, a special case of ordinary least squares regression, that is commonly used in economics in which the outcome variable is binary, and the dependent variables may be binary or continuous. We fit the linear probability model regression below for each surge discontinuity,  with \textit{Ride} indicating whether a particular request leads to a trip, and include all trips on either side of the surge discontinuity:

\begin{equation}
\label{RegDisconDes}
 Ride = \beta_0 + (\alpha * i_1 * i_2) + (\beta_1 * i_1) + (\beta_2*(1 - i_1) * i_2) + 
  (\beta_3 (1 - i_2) * x_1)  + (\beta_4 * i_2 * x_1) + \epsilon   \hfill
\end{equation}
where $\alpha$ indicates the drop in rides around a discontinuity, $i_1$ is a decision variable indicating whether the surge of that particular trip lies within 0.01 of a surge price discontinuity, $i_2$ is a decision variable that denotes whether the surge for this particular trip is to the right of the price discontinuity (\textit{i.e.} its discretized surge level is higher than the discretized surge point, thus a trip with a continuous surge level of 1.451 discretized to 1.5x surge will have a value of 1, whereas 1.449 discretized to 1.4x surge will have a value of 0), and $x_1$ is the actual (non-discretized) surge value, and the $\beta$s are the coefficients, and $\epsilon$ the error. To compute this at each jump level, we make use of trips that are on either side of the discontinuity. Since we compute this for each surge discontinuity, $\alpha$ helps capture the change in number of trips due to change in price because of the surge level. Our calculation of consumer surplus and price elasticity is taken from the approach detailed in an economics paper by \cite{cohen2016using} to compute consumer surplus for Uber across four major markets in the United States. To sum up, price elasticities are first estimated at different surge levels using the regression equation we applied in Equation~\ref{RegDisconDes}, and then we utilize those price elasticities for computing consumer surplus.


This regression discontinuity design we compute at each price discontinuity (Equation \ref{RegDisconDes}) helps estimate $\alpha$, which indicates the change in the number of trips that occur at that discontinuity due to the change in price. We can then make use of this $\alpha$ to compute price elasticities for price discontinuities as below:  


\begin{equation} 
\label{PriceElasEq}
E_{p} = \frac{\delta_Q}{\delta_P}  = \frac{\frac{\alpha}{N_p}}{\delta_P}
\end{equation} 

\noindent where $N_p$ is the proportion of trips that occur at a particular price $p$. 



At this point we run into another data-related challenge: surge levels are unavailable in ride-hailing data. To get around this, we make use of RANdom SAmple Consensus (RANSAC) regression to determine when surge pricing was in effect and what the surge levels were. When data contains outliers, RANSAC can be used for the robust estimation of model parameters from a subset of `inliers' (i.e. the data points that are not outliers) from the dataset. The intuition behind using RANSAC is as follows. According to Uber, surge pricing goes into effect when there are an unusually large number of people requesting trips at the same time \cite{Uber}. In other words, surge pricing occurs when a non-standard or much higher than normal number of customers try to book trips simultaneously. If all trips that take place at 1.0x (no surge) are standard trips, or inliers, then trips that occur at surge pricing would be considered the outliers. Once we have the surge level in effect for each ride (1.0x or higher), we can use rides on either side of a surge level in the regression continuity design to estimate price elasticities. 

\renewcommand{\algorithmicrequire}{\textbf{Input:}}
\renewcommand{\algorithmicensure}{\textbf{Output:}}       
\begin{algorithm}[hbt!]
\caption{Affordability}\label{alg:two}
\begin{algorithmic}[1]
\label{AffAlgo}
\REQUIRE Ride-hailing data ($d$) for a group
\ENSURE Total consumer surplus for the group
\STATE Compute surge level of rides: $Surge \gets RANSAC(d)$
\FOR{$\textit{surge level } s \gets 1.1x$ to $N$ }
\STATE Compute \textit{regression discontinuity design} around $s$ (Equation 3) 
\STATE Compute $ E_p \gets \frac{\frac{\alpha}{N_p}}{\delta_P}$ at $s$ (Equation \ref{PriceElasEq})
\ENDFOR
\FOR{trips at surge s $\gets 1.0x $ to $N$}
\STATE consumer surplus  $ \Delta_c \gets \sum_{i = s + 0.1x}^{N} {E_p}_i * numTrips_i * ((\frac{i - s}{s})*100) * {avgFare}_s$
\ENDFOR
\end{algorithmic}
\end{algorithm}
To reiterate, consumer surplus is the difference between the price a consumer is willing to pay for a product or service, and the price they are actually charged. Algorithm \ref{AffAlgo} details its sketch. Specifically, we look at ride-hailing data for a group (EDA or non-EDA residents), and first estimate trips' surge levels (line 1). Next, we compute a regression discontinuity design around each successive surge level, and then estimate price elasticity at that surge level (lines 2-5). When considering trips that take place at 1.0x surge (i.e., no surge), we take the price elasticity at the next surge level, 1.1x, and calculate the number of trips that would have happened if the customers shown 1.0x surge had been shown a 1.1x surge instead. We then multiply this number of potential trips with the difference in fare (which is 10\% in this case) and the average price actually paid at 1.0x surge. This is the surplus for up to surge 1.1x for those riders that were charged 1.0x surge. We then replicate this for each pair of successive surge levels, and sum it all up to arrive at the total consumer surplus estimate (lines 6-8). Finally, Algorithm \ref{AffAlgo} outputs consumer surplus for trips for the group.
\section{Fair Pricing Mechanisms}
With the tailored fairness metrics for ride-hailing, we now detail fair pricing mechanisms to help reduce disparate impact on disadvantaged communities. We focus our efforts on pricing EDA trips to help address the disparate impact on EDA residents which can help the City of Chicago in addressing concerns about mobility inequity for its residents \cite{OnTo2050}. In addition, to allow for flexibility when accounting for different situations, we define two variants of our pricing mechanism: variable discounting and fixed discounting. With variable discounting, the discount given to EDA-trips may vary depending on many factors, such as time of day, demand, etc., while all EDA-trips receive the same discount (e.g., 15\%) with the alternative fixed discounting. Such a dual approach reflects the needs of the platform and/or the government's policy-making. Specifically, the platform might require the fixed discounting route because that may be easier and more straightforward to implement; on the other hand, the government may think it is justified to give different trips different discounts depending on the conditions at the time, and the increased complexity that may come with implementing such a system is not a major concern. 


\subsection{Variable discounting: FairRide}
We first propose a new pricing mechanism called FairRide. The intuition behind FairRide is that a pricing mechanism should take different riders' differing ability to afford rides into account in order to reduce disparate impact on the most disadvantaged in society. 
To that effect, we propose a mechanism which prices rides for EDA-regions separately. We focus on EDA-trips, and run a multiple regression model to determine pricing only for rides that begin or end in an EDA (EDA-trips). Considering only trips by EDA residents leads to rides to be priced in accordance with the riders' ability to afford them. In Section 4 we compare FairRide with machine learning models commonly employed for pricing in the literature 
\cite{gu2020empirical,rathan2019crypto,miao2017influential,mohd2020overview,wolk2020advanced,alkhatib2013stock}, and find that looking only at EDA-trips leads to most models resulting in more trips for EDA residents than the current pricing mechanism, and lead to a higher relative rideability ($R^2$), and FairRide outperforms them all. A naive baseline of simply applying a \$5 discount on all rides is also implemented. 


\subsection{Fixed discounting: FixedFairRide}
The second fair pricing mechanism, called \textit{FixedFairRide}, solves an optimization problem to maximize EDA-trips while maintaining or increasing revenue in a manner that offers a consistent, fixed, discount ($\delta$) to all EDA-trips. In other words, we determine a fixed amount to discount EDA-trips by to ensure a fixed discounting policy. Our overarching goal is to determine the optimal value, $\delta$, for discounting EDA-trips such that relative rideability increases and the number of EDA trips is maximized. This can be mathematically formulated as: 

\begin{equation}
   \eta = \max_{\delta} \sum_t \left[ \sum_{l} N (\delta * p) \right]
\end{equation}
\vspace{0.02cm}

\noindent where $\eta$ represents the total number of rides, $t$ denotes a particular time period, $l$ represents EDA location pairs, $\delta$ denotes the discount for EDA-trips, $p$ is the price of the EDA trip, and $N$ is the number of EDA trips at that price.

Now we introduce two constraints to this optimization according to two possible scenarios. In the first setting, the government covers the discount for EDA residents as a subsidy. Thus, the platform and the driver receive the same amount of revenue they receive currently from a ride, but the EDA resident would get a discount from the original price that will be topped up by the government subsidy. Here we do not need to consider revenue, since the price charged by the platform does not change. Nor do we need to consider the feasibility of the trip for the driver, since the driver still receives the same exact fare as the pre-discount price. However, the government may wish to set a ceiling for the discount they offer (say, e.g., 30\%), which can be a constraint for this setting:

\begin{equation}
    0 < \delta < n
\end{equation}

\noindent where $n$ is the maximum discount the government is willing to subsidize.

In the second setting, we assume the platform offers to cover the discount themselves to help address the disparate impact. However, the platform would likely prefer the total revenue to not drop. Thus, the alternative constraint is:
\begin{equation}
\label{rev_1}
    Revenue \geq r,  
\end{equation}
\begin{equation}
    0 < \delta < 1
\end{equation}

\noindent where $r$ denotes the total revenue by current pricing, while $\delta$ must be greater than 0 indicating the discount for EDA-trips is always positive. 
In addition, we observe that we can increase revenue beyond $r$ simply by adding a tremendous amount of new, heavily discounted trips. However, below a certain price, a trip may not be worth it for the driver since they may not earn enough from it to cover their costs and make a profit. To consider the feasibility of the trip for the driver, we enforce another floor for revenue so that the driver is not in danger of not earning anything from the trip and the floored constraint thus becomes:

\begin{equation}
\label{rev_2}
    Revenue \geq max(r,\  \eta *p_{min})
\end{equation}
where $r$ is the total revenue with current pricing, $\eta$ the total number of rides. $p_{min}$ is the minimum price for a trip which can be a function of factors like travel time, distance, demand, etc, deemed important by the platform for a trip. 

The discontinuities in the price due to surge levels make this a non-convex problem, and not straightforward to solve. We therefore employ grid search \cite{lerman1980fitting} while varying values for $\delta$. 

\section{Experimental Evaluation}


In this section, we use the real-world data for the City of Chicago to run experiments. We use this dataset since this is, as far as we are aware, the only publicly available dataset that contains price (or fare) information for trips, which is necessary to address bias in pricing in ride-hailing~\cite{li2021time}. The data from all ride-hailing platforms that operate within the city of Chicago are aggregated, standardized, and anonymized before release. Which ride-hailing platform serviced a trip is not identified. For each trip, the dataset provides the trip start (or `pick-up') time, trip end (or `drop-off') time, each rounded off to the nearest 15 minutes; trip pick-up and drop-off locations at the level of Chicago census tract or community area; duration of the trip in seconds; and the fare of trip, rounded to the nearest \$2.50. Locations for pick-ups or drop-offs outside the city limits of Chicago are unavailable. Drivers and riders are not identified. We look at trips from January to October 2021, and focus our attention on trips that were not authorized as `shared.' Since the Covid-19 pandemic began in early 2020, most ride-hailing services disabled the option for shared rides. 

Finally, we perform a spatial-join between the pick-up and drop-off coordinates of this dataset with a spatial dataset released by the City of Chicago that identifies EDA regions in Chicago, to help identify trips within our data that begin or end in EDA- or non-EDA-regions. 


\subsection{The Profound Bias}

We use the metrics we proposed in Section 4 to unveil the intrinsic discrimination of the ride-hailing companies' current pricing mechanisms. We find that $R^2 = 0.33$. Such a low $R^2$ indicates the pervasiveness of real bias, and the extent of disparate impact it could cause on disadvantaged communities. We also observe below a stark difference in consumer surplus and affordability. 



\begin{figure}[!htb]
     \centering
     \begin{subfigure}[b]{0.49\textwidth}
         \centering
         \includegraphics[width=\textwidth]{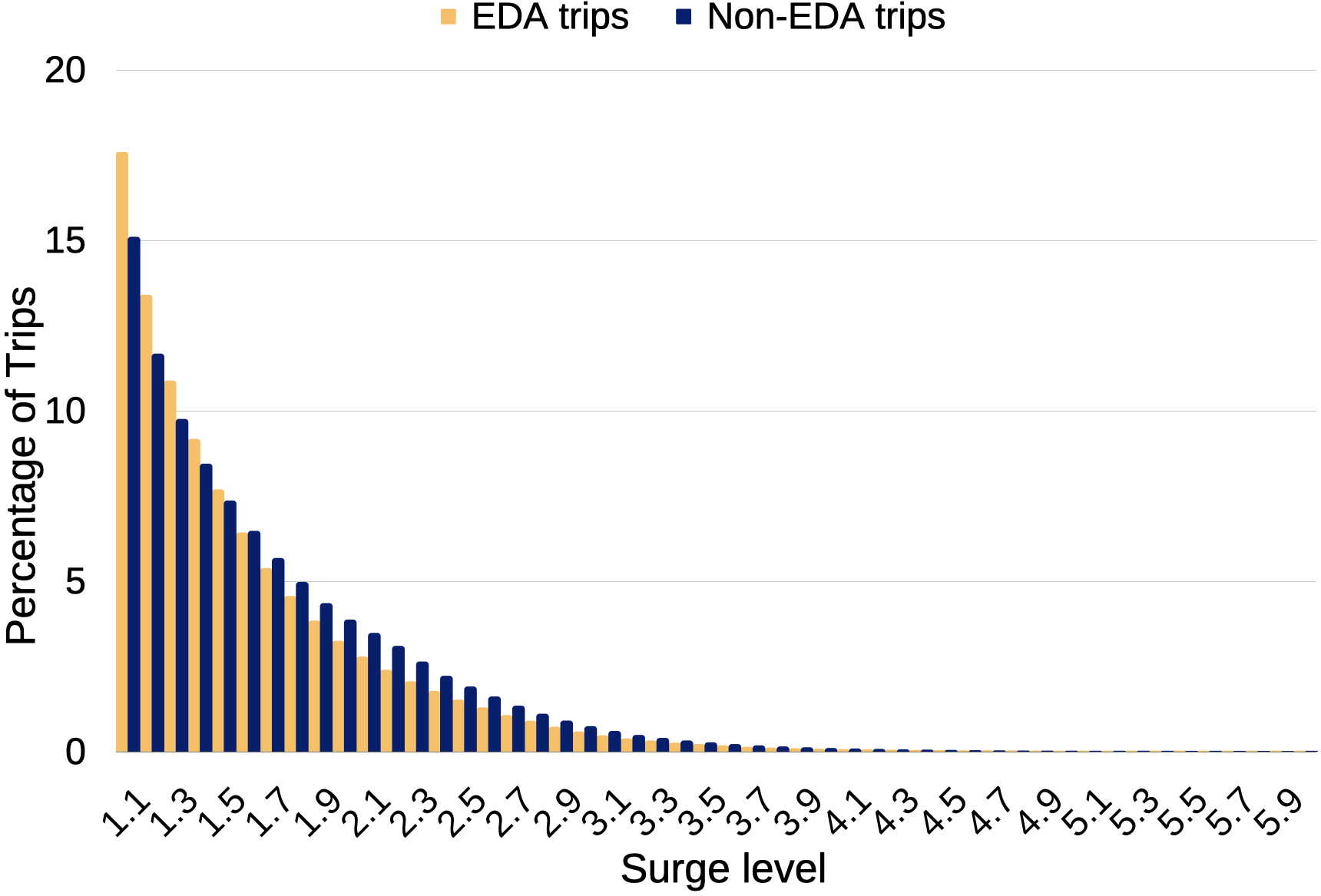}
         \caption{Share of trips at surge levels for the first ten months of 2021.}
         \label{fig:surge_counts}
     \end{subfigure}
     \hfill
     \begin{subfigure}[b]{0.49\textwidth}
         \centering
         \includegraphics[width=\textwidth]{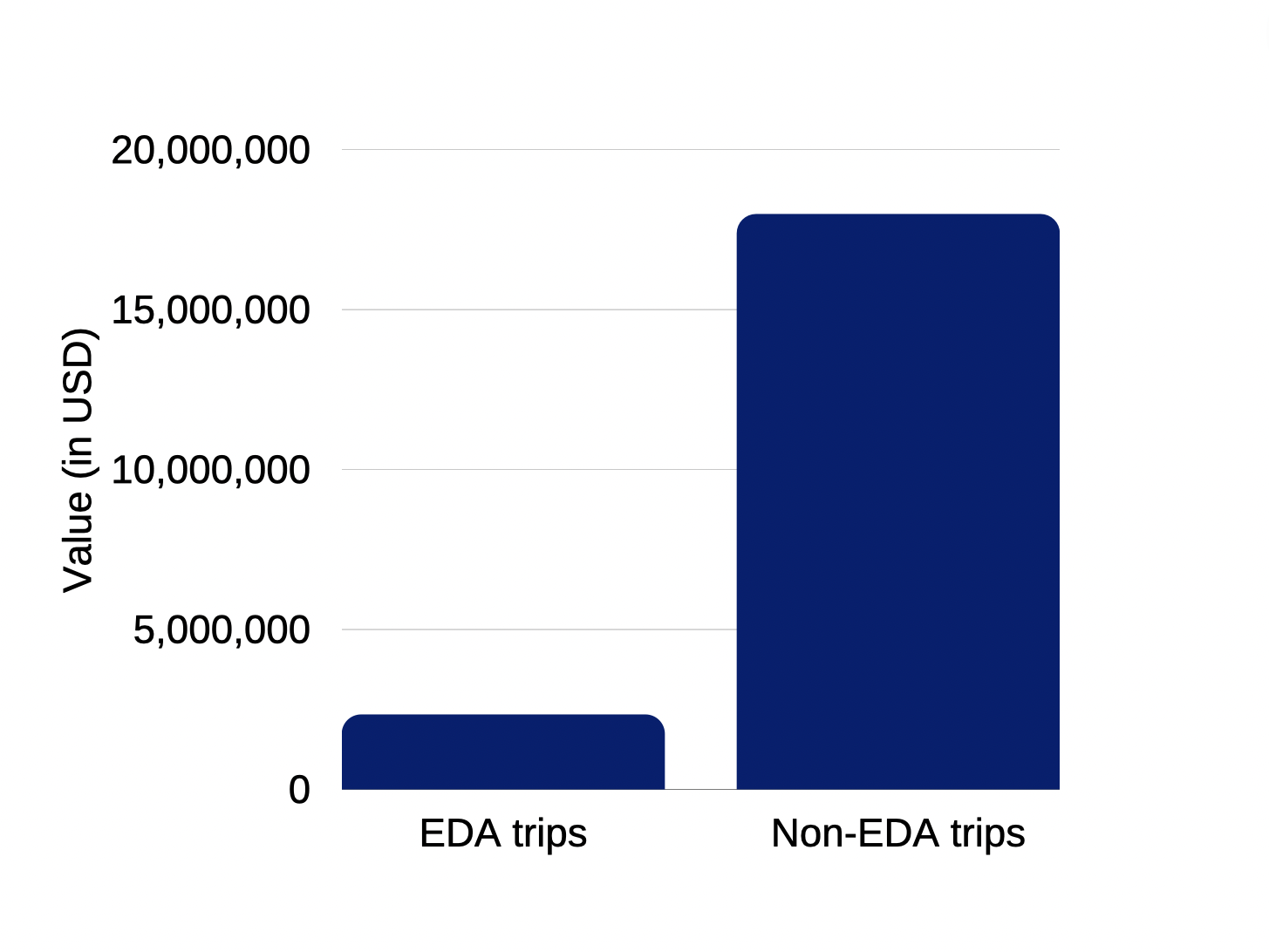}
         \caption{Consumer Surplus for trips.}
         \label{fig:surplus}
     \end{subfigure}
     \caption{Surge levels and affordability ($\Delta_c$).}
     \hfill 
\end{figure}

Next we look at affordability via consumer surplus.  
We first use RANSAC regression to estimate when surge pricing was in effect, and the surge levels for rides. The general trend can be observed in Figure \ref{fig:surge_counts}: generally, as the surge multiplier increases, the number of trips that occur decrease. While we do not have ground truth for surge levels to measure how accurate the predictions from our model are, the trend we observe is in line with the surge level trends in \cite{cohen2016using}, who had access to ground truth as the work was in conjunction with Uber. 
We can also make the following observations from Figure \ref{fig:surge_counts}. First, a greater share of EDA-trips take place at lower surge levels as compared to non-EDA-trips. 
Second, the number of EDA-trips falls below 1,000 at surge level 4.4x, when the average trip price is \$62.06. But for non-EDA-trips, it takes until surge level 5.9x for the number of trips to fall below 1,000 trips, at an average price of \$97.02. 

\vspace{-0.2cm} 
\vspace{-0.6cm}

Now that we know the surge levels for trips, we use the equations from Section \ref{DI} to calculate price elasticities and consumer surplus. Price elasticity is negative for both EDA-trips and non-EDA-trips, except in cases when surge levels are very high ($>$ 8.0x) and with very few rides (typically $<$100). This indicates that as prices increase, demand typically decreases, and vice-versa. We find that the consumer surplus of non-EDA-trips is much higher than that of EDA-trips. Examining only surge levels with a reasonable number of rides (at the highest surge levels we observe less than a 100 rides at each level, so we do not include these in the following figures), the total consumer surplus for non-EDA-trips is \$17,972,629.829, while the surplus for EDA-trips is \$2,326,555.876 over the 10 month period we look at (Figure \ref{fig:surplus}). On dividing by the total number of rides serviced in each category, we get an average consumer surplus of \$67.11 for non-EDA-trips, and an average consumer surplus of \$36.76 for EDA-trips. 

In 2016, \cite{cohen2016using} estimated that the consumer surplus for UberX across the entire United States in the year 2015 was \$6.8 billion, thus one day's consumer surplus across all cities in the US by their estimates would be \$18 million. Further, these calculations are for 7 years ago, and they would likely be higher now. Compared to these figures, and considering that Chicago is a major city in the United States, our estimates are likely lower than the true values. Thus riders beginning or ending trips in non-EDA areas are able to afford ride-hailing services far more easily than riders that begin or end trips in EDA neighborhoods.


\subsection{Pricing Mechanisms}
\subsubsection{FairRide}
We compare FairRide to machine learning models used for pricing 
\cite{gu2020empirical,rathan2019crypto,miao2017influential,mohd2020overview,wolk2020advanced,alkhatib2013stock}. We also compare against a naive baseline, which is a discount of \$5 applied to all EDA rides. 

We find that training on data organized by region leads to lower prices for approximately 35.7\% of EDA-trips, and a 35.59\% increase in rides for the period we examine in 2021 for rides beginning or ending in an EDA. In other words, as rides become more affordable, there are an additional 1,803,514 rides that begin or end in an EDA. As a result, with \textit{FairRide} Relative Rideability ($R^2$) increases by 35.6\% to 0.457, and affordability ($\Delta_c$) increases by 22.5\% to \$2,850,552.77. 
%


\vspace{-0.6cm}
\begin{table}
\caption{Number of trips, $R^2$, and affordability from different models.}
\label{NumTrips}
\centering
\begin{tabular}{ |c|c|c|c| } 
\hline
Model               & \thead{$\eta$ (\% change)} & \thead{$R^2$ (\% change)}& \thead{$\Delta_c$ (\% change)}\\ 
\hline
Original            & 5,066,849                              & 0.337                                 & 2,326,555.87 \\ 
\textbf{FairRide}   & \textbf{6,870,363}                     & \textbf{0.457 (+35.60)}           & \textbf{2,850,552.77 (+22.52)}\\
Mohd et al. (2020) \cite{mohd2020overview} & 6,867,535                              & 0.456 (+35.31)                    & 2,722,461.33  (+17.01) \\
Miao (2017) \cite{miao2017influential}   & 6,868,152                              & 0.456 (+35.31)                    & 2,818,648.50 (+21.15) \\
Wolk (2020) \cite{wolk2020advanced}     & 6,864,923                              & 0.456 (+35.31)                    & 2,534,351.94 (+8.93) \\
Baseline (-\$5)                         & 6,618,494                              & 0.440 (+30.56)                    & 2,739,784.93 (+17.76) \\
Gu et al. (2020) \cite{gu2020empirical} & 6,843,205                              & 0.455 (+35.01)                    & 2,612,239.28 (+12.27) \\
Rathan et al. (2019) \cite{rathan2019crypto} & 6,385,843                              & 0.424 (+25.81)                     & 2,667,079.84 (+14.63) \\
Alkhatib et al. (2013) \cite{alkhatib2013stock}& 6,608,789                              & 0.439 (+30.26)                    & 2,476,128.68 (+6.42) \\
\thead{\textbf{FixedFairRide}($p_{min} = 12$)} & \textbf{8,142,519} & \textbf{0.541 (+60.53)} & \textbf{3,201,963.73 (+37.62)}\\
 \hline
\end{tabular}
\end{table}
\vspace{-0.72cm}

\subsubsection{FixedFairRide}
We then test \textit{FixedFairRide} with different values of $p_{min}$ (the minimum price for a trip) and observe that as $p_{min}$ increases, the value of $\delta$ (the trip discount) decreases (Table \ref{Discount}). 

Thus, as the minimum price per trip increases, the lower the discount will be. Each of these values of $p_{min}$ results in more EDA-trips than the current pricing
\begin{table}[h]
\caption{Discount $\delta$ different minimum trip price $p_{min}$.}
\label{Discount}
\centering
\begin{tabular}{ |c|c c c c c c c c c c c| } 
 \hline

$p_{min}$ & 5 & 6 & 7 & 8 & 9 & 10 & 11 & 12 & 13 & 14 & 15 \\
\hline
$\delta$ & 0.77 & 0.72 & 0.67 & 0.62 & 0.57 & 0.52 & 0.47 & 0.42 & 0.36 & 0.30 & 0.24 \\
 \hline
\end{tabular}
\end{table}

\noindent mechanism, all also lead to more trips than \textit{FairRide} up till $p_{min} = 15$, which results in 6,824,375 trips. There is a significant increase in Relative Rideability ($R^2$) as well as affordability ($\Delta_c$ or consumer surplus) for EDA residents. In the dataset we use for our experiments the lowest fare value is \$2.50, and the highest proportion of trips (approximately 14\%) happen around \$10. Thus it is likely a minimum trip price much higher than \$10 will not be practical in the real world. 




\section{Conclusion}
Although a number of studies in the recent past have explored pricing for ride-hailing services, including looking at fair compensation for drivers, and insurance policy as a safeguard against ride-hailing using personal information for pricing, none have looked at fairer pricing to reduce disparate impact on disadvantaged residents. We proposed applicable fairness metrics to unveil the intrinsic bias in ride-hailing along with a flexible pricing mechanism to price rides more fairly and make trips more affordable for disadvantaged residents who are under-served by public transportation. The empirical experiments reveal 
the profound bias caused by existing ride-hailing pricing, and show pricing trips by our mechanism leads to more affordable and equitable ride-hailing services which could assist government policy-making for fair ride-hailing policies. 


%
%
%
\bibliographystyle{splncs04}
\bibliography{mybibliography}
%




\end{document}